# Optofluidic light routing via analytically configuring streamlines of micro-flow

Ruopeng Yan[1], Yunran Yang[1], Xin Tu[1], Tianye Huang[1], Y. Liu (刘泱杰) [2], Chaolong Song[1],*

1. School of Mechanical Engineering and Electronic Information, China University of Geosciences, Wuhan, China
2. School of Physics and Electronic Science, Hubei University, Wuhan 430062, Hubei Province, P. R. China

*corresponding author songcl@cug.edu.cn

## ABSTRACT

Transformation optics (TO) is a new method to design metamaterials that can manipulate electromagnetic fields. Inspired by the traditional TO techniques which is mostly based on the solid metamaterials with a limited range of tunability, a novel streamline tracing-based transformation optofluidics (STTOF) method is proposed to manipulate the light path by analytically designating the light-carrying streamlines of the flow in a two-dimensional circular bounded domain. A dipole flow model is built to analytically calculate the streamlines of the flow field inside the domain which allocates the optical/fluidic source and sink pairs at arbitrary positions. Liquid core/liquid cladding ($L^2$) configuration is used in the experiment to trace the light via a specific streamline. Experimental results verify that the light paths agree well with the theoretical predictions, and demonstrate that a good range of tunability can be achieved by adjusting the flow rates and the source-sink positions of optical/fluidic source and sink pairs.

# 1 Introduction

In the recent years, transformation optics (TO) has become an increasingly popular tool for designing of electromagnetic media with various spatial distributions of permittivity and permeability (metamaterial), which offers great convenience for electromagnetic wave manipulating (Chen et al. 2010; Ergin et al. 2010; Leonhardt 2006; Pendry et al. 2006; Li and Pendry 2008; Pendry et al. 2012; Vakil and Engheta 2011). One of the most important applications is to conceal object from visible light, which is generally referred to as cloaking of objects (Chen et al. 2011; Li and Pendry 2008; Liu et al. 2009; Ni et al. 2015; Schurig et al. 2006; Valentine et al. 2009; Zhu et al. 2017). Currently, most of TO methods involves using metamaterials with a wide range of permittivity and permeability, which enable their spatial variation via construction of different nano-structures of materials. However, metamaterials are mostly solid-based which leaves no room for flexibility once after fabrication. As a result, the light path therein is often fixed in a given design of metamaterial.

Meanwhile optofluidic devices have also attracted wide attentions from research communities for their potentials in optical tunability and novel functionalities.(Liang et al. 2018; Shi et al. 2018a; Shi et al. 2017; Shi et al. 2018b; Song et al. 2017; Song and Tan 2017). A number of optically tunable devices have been explored using fluidic configurations, such as routers (Esseling et al. 2015; Muller et al. 2014), optical switches (Huang et al. 2012; Lim et al. 2011), reflective index (RI) sensors and biosensors(Fan and White 2011; Zhang et al. 2018). On one hand, smooth fluid boundary as an ideal optical interface can be dynamically controlled by changing the flow field, thereby to conveniently manipulate the light path. On the other hand, fluid diffusion provides concentration distribution to generate specified RI profile. Recently, many efforts have been devoted to design a specified flow field to map out a route for light guiding (Odeh et al. 2017; Schmidt and Hawkins 2008; Shi et al. 2015), focusing(Song et al. 2009a; Song et al. 2009b; Song et al. 2009c), splitting(Cho et al. 2009; Nguyen et al. 2007; Yang et al. 2012b) and bending (Liu et al. 2017; Yang et al. 2012a; Yang et al. 2011). Of them especially transformation optofluidics have been applied to manipulate light beam using gradient index (GRIN) distribution achieved by the convection-diffusion(Liu et al. 2017; Yang et al. 2012b; Yang et al. 2011). However, the flow field mainly depends on the geometry shape of fluid boundaries and strengths of driven forces. Thus generally computational fluid dynamics (CFD) methods are required calculate the concentration distribution ad hoc case by case, which increases the complexity and difficulty for the entire design process. Therefore a guiding rule of thumb will be highly welcome to ease the design process.

In this work, we propose a novel streamlines-tracing-based transformation optofluidics (STTOF) method. Unlike TO method conducting space coordinate transformation, STTOF realizes an optical transformation by switching the coordinate of streamlines which is closely related to boundary conditions like geometrical boundary, flow rate etc. To forecast the light propagation along the flow, a theoretical model is firstly established based on multipole flow theory to analytically trace out the streamlines of the flow inside the domain. As the light can be designed to transport on the fluid flow along the same path using liquid core/liquid cladding ($L^2$) configuration, the light path can be manipulated via designating a specified streamline for the fluid flow (Lee et al. 2012; Tang et al. 2008; Wolfe et al. 2004). The inlet/outlet on the boundary for light and fluid serves as optical and fluidic source/sink respectively (Fig. 1(a)-(c)). Streamlines with symmetric and asymmetric geometries, enabled by setting source/sink pairs at various locations, are both investigated theoretically and experimentally. Optical performance of the device is also studied by measuring the transmission loss when the light beam transmits and bends.

## 2 Theoretical analysis

The proposed STTOF is a method aimed at manipulating the light propagation by purposely devising the flow field within the domain. The L$^2$ configuration enabled by microfluidic manipulation has been widely used as the basic structure for light wave guiding, since light can be transported and maneuvered with the mass transportation(Fan et al. 2016; Levy et al. 2006; Lim et al. 2008). In this way, one can dictate the light path by driving the liquid core to selectively follow a specific streamline. Therefore, the pre-requisite is to analytically predefined the streamlines. For planar structures of microfluidic architectures, multipole flow model can be employed to calculate diversiform streamline distributions under a specific boundary condition with designated positions of sources/sinks pairs. In this work, we analytically and experimentally study the manipulation of light path via routing the streamlines in a circularly bounded domain.

To calculate the streamlines of two-dimensional dipole flow in a circular domain, a source-sink pair model is assumed to be arbitrarily placed at S$^+$(-$x_0,y_0$) and S$^-$($x_0,y_0$) on the edge of a circular domain with a radius of $R$ (Koplik et al. 1994; Song et al. 2009c) . The complex potential of the flow field can be described as:

$$W = \phi + i\psi = \ln \frac{(z+x_0 - i \cdot y_0)\left(z + \frac{R^2}{x_0 + i \cdot y_0}\right)}{(z-x_0 - i \cdot y_0)\left(z + \frac{R^2}{-x_0 + i \cdot y_0}\right)} \quad (1)$$

Where $z=x+iy$, the real part $\Phi$ denotes the velocity potential and the imaginary part $\Psi$ represents streamline function. The streamlines are equipotential lines of stream function $\Psi$, as shown in Fig. 1 (g)-(i). For simplicity, neglecting the lower half plane ($y<0$), the streamline function $\Psi$ can be:

$$\psi = 2arctan\frac{y-y_0}{x-x_0} + 2arctan\frac{y-y_0}{x+x_0} \quad (2)$$

The streamlines are a group of curves($\Psi=C_i$), and the moving point $P(x,y)$ on the curves should obey the following equation:

$$arctan\frac{y-y_0}{x-x_0} + arctan\frac{y-y_0}{x+x_0} = C_i (i=1,2,3\cdots) \quad (3)$$

The two terms on the left side of equation (3) represents the angel $\alpha$ and $\beta$ respectively, depicted in Fig. 2(a). On the right side, $C_i$ indicated a cluster of constants. When $C_i$ and the coordinates of source-sink pair are given, moving point $P$ located on an exclusive arc with radius of $r=x_0/sin(\alpha+\beta)$ . Thus, with a given position, each streamline can be well defined from Eq. (3).

To align the light path with the streamline, L$^2$ configuration was [I do not know but perhaps you could use past-tense throughout the maintext. ] utilized to form a structure that core liquid is sandwiched by two cladding flows, shown in Fig. 1(d)-(f). This configuration ensures that the light be restricted to propagate in the core as long as the employed core fluid has a higher RI ($n_{core}$) than that of the cladding fluid ($n_{cladding}$). Knowing that the flow rate of core is sufficiently small compared to that of cladding flow, the core flow along with the light transmission follows a specific streamline. Therefore, the routing of light path can be achieved by controlling the flow rates of the two claddings. The derivative of the complex potential $W$ [Eqn (1)] provides the velocity field $V=(u,v)$, and the real and imaginary part represents velocity component in $x$ $y$-plane, respectively. When $x=0$ the velocity on $y$-axis is as follows:

$$\begin{cases} u = \dfrac{4x_0}{R^2 + y^2 - 2y_0 \cdot y} \\ v = 0 \end{cases} \quad (4)$$

Suppose the core flow rate is much smaller than those of the two claddings, the flow rate of cladding A and Cladding B (Fig. 1 (g)-(i)) can be calculated by integrating velocity along the $y$-axis ($x=0$):

$$\begin{cases} Q_{claddingA} = \int_B^A u \cdot dy = \dfrac{4x_0}{\sqrt{R^2 - y_0^2}} \left[ \arctan\left(\dfrac{R - y_0}{\sqrt{R^2 - y_0^2}}\right) - \arctan\left(\dfrac{y_B - y_0}{\sqrt{R^2 - y_0^2}}\right) \right] \\ Q_{claddingB} = \int_C^B u \cdot dy = \dfrac{4x_0}{\sqrt{R^2 - y_0^2}} \left[ \arctan\left(\dfrac{y_B - y_0}{\sqrt{R^2 - y_0^2}}\right) - \arctan\left(\dfrac{-R - y_0}{\sqrt{R^2 - y_0^2}}\right) \right] \end{cases} \quad (5)$$

where B is the intersection point between the streamline and y-axis, $\overline{AB}$ is the distance from upper boundary to the specific streamline, and $\overline{BC}$ is the distance from streamline to lower boundary, shown in Fig. 2(b). From equation (5), the curvature of the streamline is related to the flow rate ratio of $Q_{claddingA}/Q_{claddingB}$. In this way, the light path can be simultaneously manipulated by hydrodynamically changing the flow rate ratio between the two cladding flows ($Q_{claddingA}/Q_{claddingB}$).

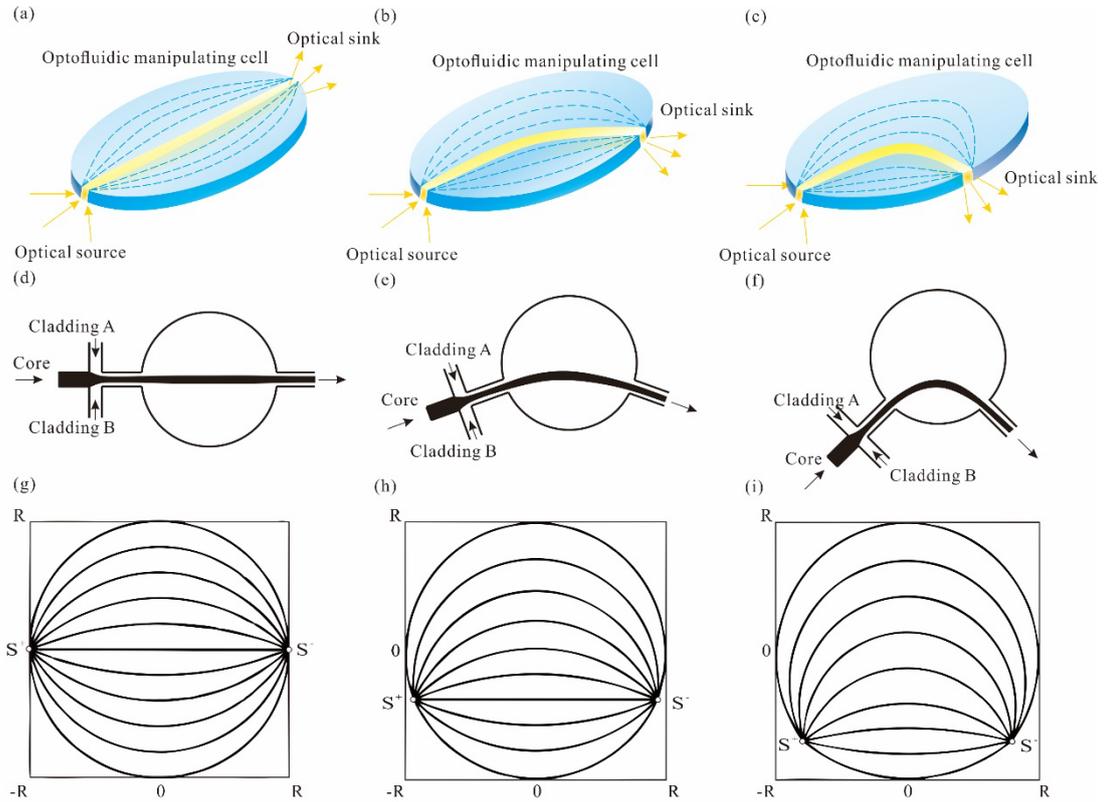

**Fig. 1.** The concept of the optofluidic optical source-sink pair with a circular optofluidic manipulating cell. Optical source-sink pairs are located at (a) 180-degree arc (b) 135-degree arc (c) 90-degree arc. The corresponding inlets and outlets of L$^2$ configuration are respectively set at (d) 180-degree arc (e) 135-degree arc (f) 90-degree arc. The yellow arrows represent the light input and output. The dotted blue lines represent the streamlines. The analytic streamlines in a circular boundary domain are shown in (g) (h) (i).

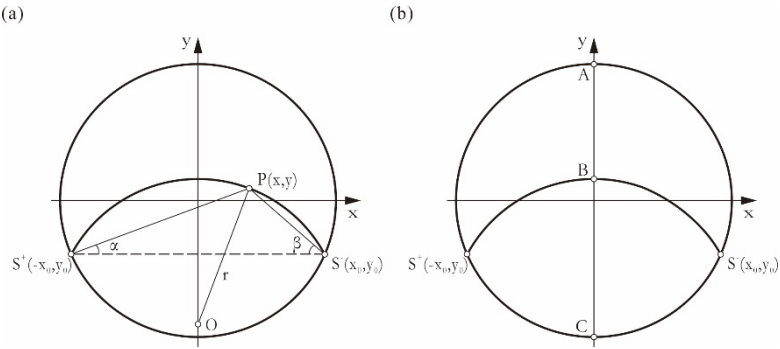

**Fig. 2.** (a) Track of moving point P satisfying Eqn (3); (b) position of a specific streamline.

## 3 Experimental materials and methods

### 3.1 Device design and fabrication

The chip design is shown in Fig.3 (a). The device consists of a circular cell with a diameter of 2mm, and six ports, served as optical and fluid source/sinks, connected to the circular cell. The width of the ports is designed to have a much smaller dimension (130um) compared to the diameter of circular cell, so that the geometry of the device can approximately satisfies the assumptions of the proposed theory.

At the source port, core inlets and cladding inlets are designed that core fluid can be sandwiched by cladding fluid to build up the $L^2$ configuration. Five sink ports are located at the margin of the circular cell to investigate the light routing with different bending radii. In between the two core inlets, a reserved channel is designed for positioning optical fiber to introduce light wave. The height of all the planar structure is set at 125μm.

The optofluidic chips were fabricated using soft lithography technique(Qin et al. 2010; Unger et al. 2000). Two components of PDMS were mixed with a weight ratio of 1:10, after which the mixture was poured onto a customized silicon wafer. Then the entirety was put into an incubator at 70°C for 1 hour. After solidification the PDMS replica was peeled off from the wafer, and punched with 0.75mm diameter wholes at each inlet and outlet. Those access wholes were connected with plastic pipes with outer diameter of 0.8 mm and inner diameter of 0.5mm for feeding of fluid. The replica was bonded on another reserved PDMS chunk with smooth surface after treating with oxygen plasma to both patterned surface and smooth surface and subsequently heating at 80°C for an hour.

### 3.2 Experiment setup

Fig.3 (b) shows the schematic of experimental setup. In the experiment, four syringe pumps (LEAD FLUID TYD02-02) were employed for liquid injection which can provide adjustable flow rates. Two of them are for cladding flow and the other two are for core flow. A beaker was connected to the outlets to collect waste. Specifically, the ethylene glycol with RI $n_{gly}$=1.432 and viscosity $\eta_{gly}$=15.5×10$^{-3}$ Pa·s at a temperature of 20°C served as cladding fluid, and cinnamaldehyde with RI $n_{cin}$=1.622 and viscosity $\eta_{cin}$=3.48×10$^{-3}$ Pa·s at a temperature of 20°C served as core fluid. Rhodamine B (Sigma-Aldrich R6626, with an absorption peak of 553 nm, an emission peak 619 nm) was mixed with the core fluid to visualize the light path. The light from semiconductor laser (Laserwave, LWGL 532-100mW-F) was coupled into a multi-mode optical fiber (THORLABS, AFS105/125Y) which was inserted into the reserved channel in the optofluidic chip. An inverted fluorescence microscope (SHUOGUANG CFM-500E) with an attached CCD was used to record the visualization of light paths, and an edgepass filter with a cut-on wavelength of 565nm was employed to avoid interference from the excitation laser.

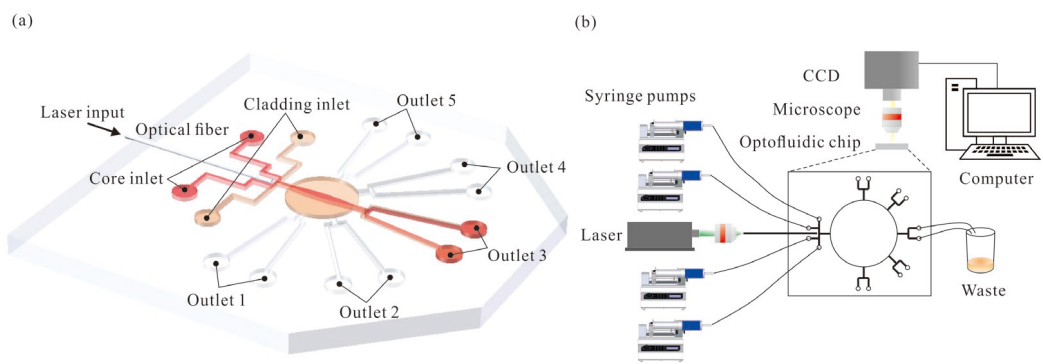

**Fig. 3.** Schematic diagrams of experimental setup: (a) optofluidic chip; (b) measurement system.

# 4 Result and discussion

Firstly, a preliminary experiment to investigate relationship between light loss and flow rate ratio $Q_{core}/Q_{cladding}$ was arranged to explore an optimized flow rate ratio between core and cladding streams. In the experiment, inlets and outlet 3 (Fig.3 (a)) were switched on, and syringe pumps were used to control the fluid injection. When both flow rate of cladding A and cladding B were kept at 100μL/min, symmetrical flow rate of cladding stream would develop a straight waveguide (depicted as Fig.1 (d)). The fluorescent intensities along the profiles at inlet and outlet were measured to estimate the light loss while adjusting the flow rate of the core stream. In each measurement, the power of input light and the exposure time of CCD were kept identical. As shown in Fig. 4, keeping the flow rate of cladding stream at 100μL/min, through continuously increasing the flow rate of core stream from 8 to 20μL/min, the sectional width of waveguide was observed to expand and the measured light transmission loss subsequently started to increase. As the streamlines in the circular domain follow arc-shapes, larger flow rate ratio $Q_{core}/Q_{cladding}$ would lead to an increased curvature of the waveguide interfaces, which could eventually result in higher optical loss when part of the light cannot satisfy the total internal reflection condition In order to maximize the optical transmission while keeping a stable formation of fluidic waveguiding, a sufficiently small flow rate of core stream was selected to perform the following experiments.

To verify the theoretical model for predicting the relationship between streamlines of the flow field and the light path, inlets and outlet 3 were switched on and the light paths along the streamlines were manipulated via tuning the flow field. According to the preliminary experiment, the flow rate of the core stream was kept at 10μL/min for a lower flow rate ratio $Q_{core}/Q_{cladding}$ and better transmission efficiency. In the experiment, the flow rate of the cladding A was increased from 20 to 80μL/min every one minute with a step of 10μL/min, and at meanwhile the flow rate of the cladding B was decreased from 80 to 20μL/min(as depicted in Fig.5 (b)). The tunable flow rate ratio between the two claddings enforced the core stream to travel with different paths which also drags the light attached to the flow. Rhodamine B was mixed into the core fluid which can be used to visualize the light path via capturing fluorescence images (Fig.5 (a)). Since all streamlines in the circular cell follow arc-shape paths (depicted as Fig. 1 (g)-(i)), experimental results show that increased flow rate ratio between the two cladding flows leads to a light path with larger bending radius. The measured bending curvature of the light path was consistent with the theoretical calculation (Eqn (5)), as illustrated in Fig.5 (c), indicating that the streamlines as well as the light path can be well predicted by the proposed theory. Moreover, the tuning range of the light path was observed to sweep across the entire domain via setting proper flow rates, which shows a potential to enhance the flexibility in light path routing compared with using traditional solid-state-based metamaterial.

In principle, the positions of source and sink can be arbitrarily allocated at the boundary of the domain with the flow field still being analytically resolved. In such cases, we also tested our device to verify that the light can travel along the predicted streamlines in the conditions of source and sink pairs at various positions. In particular, three positions for source-sink pairs were tested in our experiments, which were located on the circular boundary by taking up 180-degree arc, 135-degree arc and 90-degree arc as illustrated in Fig. 1 (d)-(f), respectively. Before the test, preliminary experiment was carried to investigate the transmission efficiency of waveguiding when the light path following different streamlines. In the cases of 45 and 90 degrees' bending of light paths, the transmission efficiencies were measured while tuning the flow rate ratio between the two cladding flows shown in Fig. 6 (a) and (b), respectively. It was found that the transmission efficiency can achieve a maximum when the light path has a tangential interconnection with the flow in the inlets and outlets channels. Therefore, to ensure light transmission with low loss, the streamline, which carries light in the circular domain, was planned to have a tangential interconnection with those in inlet/outlet channels by working out the flow rates using eqn. (5) for experimental study. In the experiment, one of the five outlets was selectively switched on each time (Fig. 3 (a)), and the others were blocked. When outlet 3 was open, core flow rate at 10μL/min and two cladding flow rates were set at 100μL/min, a straight light path was realized, as illustrated in Fig.7 (a). When outlet 2 was open, core flow rate was set at 10μL/min, cladding A flow rate at 20μL/min and cladding B flow rate at 18μL/min, a beam with bending angle 45° was realized as shown in Fig.7 (c). When outlet 1 was open, core flow rate was set at 10μL/min, cladding A flow rate at 20μL/min and cladding B flow rate at 18μL/min, a beam with bending angle 90°was achieved as shown in Fig.7 (e). The experimental results indicate that point-to-point light transmission can be manipulated by positioning the corresponding source and sink pairs which, in principle, can be distributed at arbitrary positions on the edge of the domain. In each case, the fluorescent intensity profile at inlet and outlet was also measured and fitted with Gaussian curves shown in Fig.7 (b), (d), (f). Owing to increased bending angle of the light path, more bending loss makes more light loss which reflects on the fluorescent intensity between the inlet and outlet.

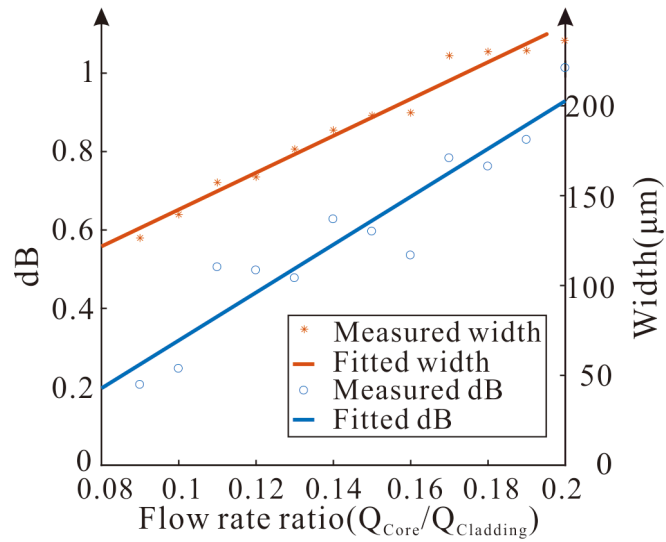

**Fig. 4.** Relationship among flow rate ratio ($Q_{core}/Q_{cladding}$), light loss, width between upper and lower interfaces in the case of inlets and outlet 3 are both open. When the flow rate of cladding stream kept at 100μL/min and the flow rate of core stream increase from 8 to 20μL/min, the width between upper and lower interfaces increases from 48μm to 220μm (see the blue line and the data points). The light loss raising from 0.58 to 1.08(dB) with the raising core flow rate (see the orange line and the data points).

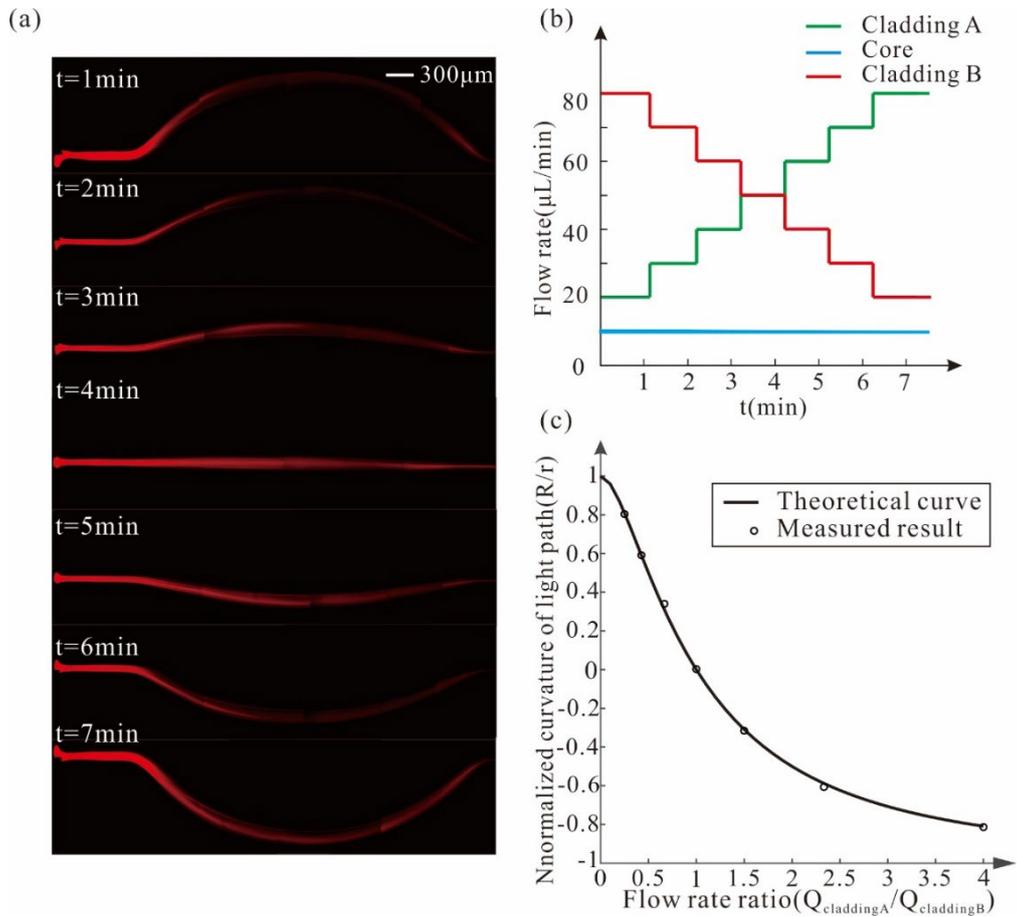

**Fig. 5.** (a) Fluorescent images of light path in 0-7min. (b) Depiction about input flow rate of core and cladding. The flow rate of the cladding A increases from 20 to 80μL/min every one minute with a step of 10μL/min. The flow rate of the cladding B decreases from 80 to 20μL/min and the core flow rate remains 10μL/min. (c) The relationship between light path and flow rate ratio $Q_{cladding\ A}/Q_{cladding\ B}$ for both theoretical prediction and measured results.

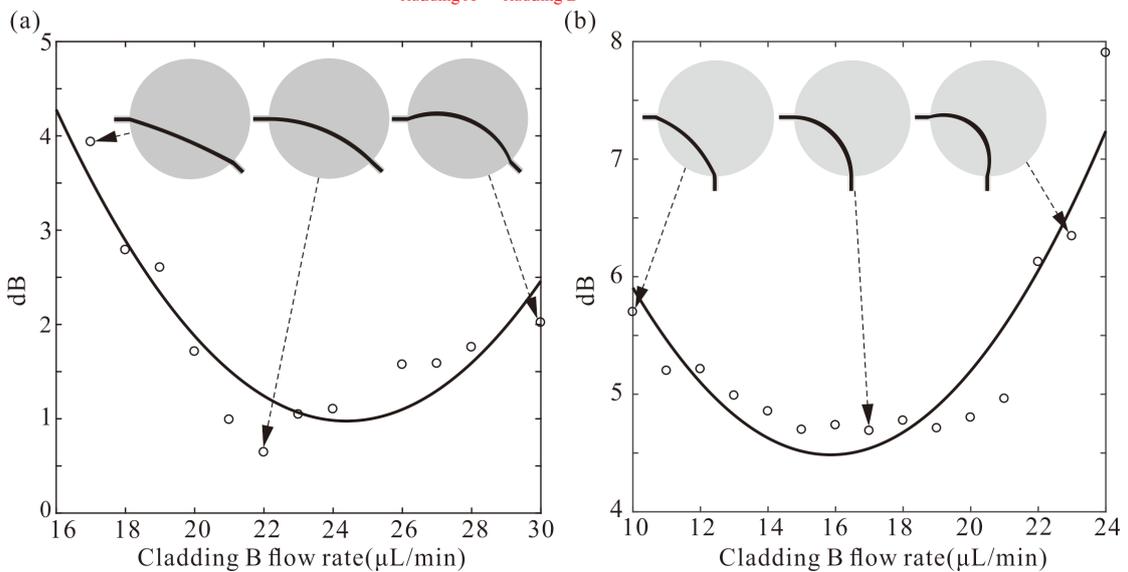

**Fig. 6.** (a) (b) Measurements of transmission efficiencies in cases of 45 and 90 degrees' bending of light. The core flow rate remains 10μL/min and flow rate of cladding A is kept at 20μL/min. Here the lines are the theoretical predictions and the data points are the experimental results. The insertion on the top are shown the process of light path variation.

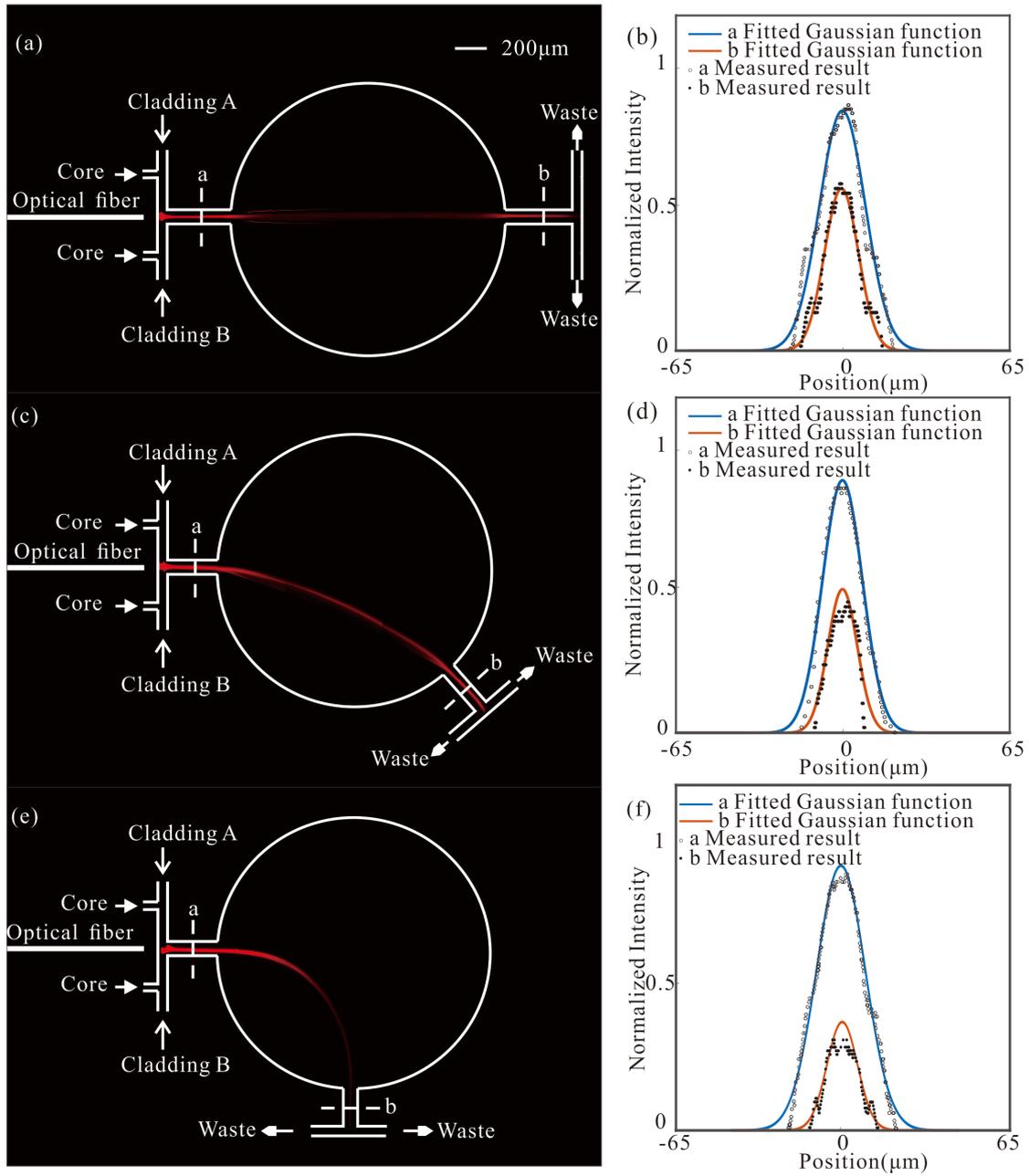

**Fig. 7.** Fluorescent images of light path in three conditions where source-sink pairs were respectively allocated at tow side of (a) 180-degree arc, (c) 135-degree arc (e) 90-degree arc. The white line indicated the geometric boundary of micro-channel wall. (a) Core flow rate at 10μL/min cladding flow rate at 100μL/min. (b) shows the fluorescent intensity profile along the dashed white lines in (a). (c) Core flow rate at 10μL/min, cladding A flow rate at 20μL/min and cladding B flow rate at 18μL/min. (d) shows fluorescent intensity profiles along the dashed white lines in (c). (e) Core flow rate at 10μL/min, cladding A flow rate at 20μL/min and cladding B flow rate at 15μL/min. (f) shows the fluorescent intensity profile along the dashed white lines in (e).

## 5 Conclusion

In conclusion, we reported a TOF method to manipulate the light path in a given fluidic domain, in which dipole flow theory was utilized to analytically pinpoint the flow field of the domain. The experiments demonstrated that the light transmission in an $L^2$ configuration follows well with the theoretically predicted streamlines. The light routing was demonstrated to be controllable via tuning the flow rate ratio between the two cladding flows. Setting fluidic/optical source and sink pair at different positions on the edge of the fluidic domain offers an extra tunable parameter for light routing. Though experimental demonstration was carried out in a circularly bounded domain, we envision that this method can also extend to light routing in domains of other geometric shapes through working out the streamline function of the flow field using superposition of various basal potential flows (Young et al. 2016; Yukselen 1996).

**Acknowledgements** We are grateful for the financial support from National Natural Science Foundation of China (61804138), Hubei Provincial Natural Science Foundation of China (2017CFB193), Wuhan Science and Technology Bureau (2018010401011297), Fundamental Research Funds for the Central Universities, China University of Geosciences (Wuhan) (CUG170608).

## References

Chen HY, Chan CT, Sheng P (2010) Transformation optics and metamaterials Nat Mater 9:387-396 doi:10.1038/nmat2743

Chen XZ, Luo Y, Zhang JJ, Jiang K, Pendry JB, Zhang SA (2011) Macroscopic invisibility cloaking of visible light Nat Commun 2:6 doi:10.1038/ncomms1176

Cho SH, Godin J, Lo Y-H (2009) Optofluidic Waveguides in Teflon AF-Coated PDMS Microfluidic Channels Ieee Photonics Technology Letters 21:1057-1059 doi:10.1109/lpt.2009.2022276

Ergin T, Stenger N, Brenner P, Pendry JB, Wegener M (2010) Three-Dimensional Invisibility Cloak at Optical Wavelengths Science 328:337-339 doi:10.1126/science.1186351

Esseling M, Zaltron A, Horn W, Denz C (2015) Optofluidic droplet router Laser Photon Rev 9:98-104 doi:10.1002/lpor.201400133

Fan S-K, Lee H-P, Chien C-C, Lu Y-W, Chiu Y, Lin F-Y (2016) Reconfigurable liquid-core/liquid-cladding optical waveguides with dielectrophoresis-driven virtual microchannels on an electromicrofluidic platform Lab on a Chip 16:847-854 doi:10.1039/c5lc01233c

Fan XD, White IM (2011) Optofluidic microsystems for chemical and biological analysis Nat Photonics 5:591-597 doi:10.1038/nphoton.2011.206

Huang PH, Lapsley MI, Ahmed D, Chen YC, Wang L, Huang TJ (2012) A single-layer, planar, optofluidic switch powered by acoustically driven, oscillating microbubbles Appl Phys Lett 101:4 doi:10.1063/1.4742864

Koplik J, Redner S, Hinch EJ (1994) Tracer dispersion in planar multipole flows Physical Review E (Statistical Physics, Plasmas, Fluids, and Related Interdisciplinary Topics) 50:4650-4671 doi:10.1103/PhysRevE.50.4650

Lee KS, Yoon SY, Lee KH, Kim SB, Sung HJ, Kim SS (2012) Optofluidic particle manipulation in a liquid-core/liquid-cladding waveguide Optics Express 20:17348-17358 doi:10.1364/oe.20.017348

Leonhardt U (2006) Optical conformal mapping Science 312:1777-1780 doi:10.1126/science.1126493

Levy U, Campbell K, Groisman A, Mookherjea S, Fainman Y (2006) On-chip microfluidic tuning of an optical microring resonator Appl Phys Lett 88 doi:10.1063/1.2182111

Li JS, Pendry JB (2008) Hiding under the Carpet: A New Strategy for Cloaking Phys Rev Lett 101:4 doi:10.1103/PhysRevLett.101.203901

Liang L, Jin YX, Zhu XQ, Zhou FL, Yang Y (2018) Real-time detection and monitoring of the drug resistance of single myeloid leukemia cells by diffused total internal reflection Lab on a Chip 18:1422-1429 doi:10.1039/c8lc00088c

Lim J-M, Kim S-H, Choi J-H, Yang S-M (2008) Fluorescent liquid-core/air-cladding waveguides towards integrated optofluidic light sources Lab on a Chip 8:1580-1585 doi:10.1039/b805341c

Lim JM, Urbanski JP, Thorsen T, Yang SM (2011) Pneumatic control of a liquid-core/liquid-cladding waveguide as the basis for an optofluidic switch Appl Phys Lett 98:3 doi:10.1063/1.3535979

Liu HL, Zhu XQ, Liang L, Zhang XM, Yang Y (2017) Tunable transformation optical waveguide bends in liquid Optica 4:839-846 doi:10.1364/optica.4.000839

Liu R, Ji C, Mock JJ, Chin JY, Cui TJ, Smith DR (2009) Broadband Ground-Plane Cloak Science 323:366-369 doi:10.1126/science.1166949

Muller P, Kopp D, Llobera A, Zappe H (2014) Optofluidic router based on tunable liquid-liquid mirrors Lab on a Chip 14:737-743 doi:10.1039/c3lc51148k

Nguyen N-T, Kong T-F, Goh J-H, Low CL-N (2007) A micro optofluidic splitter and switch based on hydrodynamic spreading Journal of Micromechanics and Microengineering 17:2169-2174 doi:10.1080/0960-1317/17/11/002

Ni XJ, Wong ZJ, Mrejen M, Wang Y, Zhang X (2015) An ultrathin invisibility skin cloak for visible light Science 349:1310-1314 doi:10.1126/science.aac9411

Odeh M, Voort B, Anjum A, Paredes B, Dimas C, Dahlem MS (2017) Gradient-index optofluidic waveguide in polydimethylsiloxane Applied Optics 56:1202-1206 doi:10.1364/ao.56.001202

Pendry JB, Aubry A, Smith DR, Maier SA (2012) Transformation Optics and Subwavelength Control of Light Science 337:549-552 doi:10.1126/science.1220600

Pendry JB, Schurig D, Smith DR (2006) Controlling electromagnetic fields Science 312:1780-1782 doi:10.1126/science.1125907

Qin D, Xia Y, Whitesides GM (2010) Soft lithography for micro- and nanoscale patterning Nature Protocols 5:491-502 doi:10.1038/nprot.2009.234

Schmidt H, Hawkins AR (2008) Optofluidic waveguides: I. Concepts and implementations Microfluid Nanofluid 4:3-16 doi:10.1007/s10404-007-0199-7

Schurig D, Mock JJ, Justice BJ, Cummer SA, Pendry JB, Starr AF, Smith DR (2006) Metamaterial electromagnetic cloak at microwave frequencies Science 314:977-980 doi:10.1126/science.1133628

Shi Y, Liang L, Zhu XQ, Zhang XM, Yang Y (2015) Tunable self-imaging effect using hybrid optofluidic waveguides Lab on a Chip 15:4398-4403 doi:10.1039/c5lc01066g

Shi Y et al. (2018a) Nanometer-precision linear sorting with synchronized optofluidic dual barriers Science Advances 4 doi:10.1126/sciadv.aao0773

Shi YZ et al. (2017) High-resolution and multi-range particle separation by microscopic vibration in an optofluidic chip Lab on a Chip 17:2443-2450 doi:10.1039/c7lc00484b

Shi YZ et al. (2018b) Sculpting nanoparticle dynamics for single-bacteria-level screening and direct binding-efficiency measurement Nat Commun 9:11 doi:10.1038/s41467-018-03156-5

Song C, Nguyen N-T, Asundi AK, Low CL-N (2009a) Biconcave micro-optofluidic lens with low-refractive-index liquids Optics Letters 34:3622-3624 doi:10.1364/ol.34.003622

Song C, Nguyen N-T, Tan S-H, Asundi AK (2009b) A micro optofluidic lens with short focal length Journal of Micromechanics and Microengineering 19 doi:10.1088/0960-1317/19/8/085012

Song C, Nguyen N-T, Tan S-H, Asundi AK (2009c) Modelling and optimization of micro optofluidic lenses Lab on a Chip 9:1178-1184 doi:10.1039/b819158a

Song CL, Nguyen NT, Tan SH (2017) Toward the commercialization of optofluidics Microfluid Nanofluid 21:16 doi:10.1007/s10404-017-1978-4

Song CL, Tan SH (2017) A Perspective on the Rise of Optofluidics and the Future Micromachines 8:17 doi:10.3390/mi8050152

Tang SKY, Stan CA, Whitesides GM (2008) Dynamically reconfigurable liquid-core liquid-cladding lens in a microfluidic channel Lab on a Chip 8:395-401 doi:10.1039/b717037h

Unger MA, Chou HP, Thorsen T, Scherer A, Quake SR (2000) Monolithic microfabricated valves and pumps by multilayer soft lithography Science (New York, NY) 288:113-116 doi:10.1126/science.288.5463.113

Vakil A, Engheta N (2011) Transformation Optics Using Graphene Science 332:1291-1294 doi:10.1126/science.1202691

Valentine J, Li JS, Zentgraf T, Bartal G, Zhang X (2009) An optical cloak made of dielectrics Nat Mater 8:568-571 doi:10.1038/nmat2461

Wolfe DB et al. (2004) Dynamic control of liquid-core/liquid-cladding optical waveguides Proceedings of the National Academy of Sciences of the United States of America 101:12434-12438 doi:10.1073/pnas.0404423101

Yang Y, Chin LK, Tsai JM, Tsai DP, Zheludev NI, Liu AQ (2012a) Transformation optofluidics for large-angle light bending and tuning Lab on a Chip 12:3785-3790 doi:10.1039/c2lc40442g

Yang Y et al. (2012b) Optofluidic waveguide as a transformation optics device for lightwave bending and manipulation Nat Commun 3 doi:10.1038/ncomms1662

Yang Y, Liu AQ, Lei L, Chin LK, Ohl CD, Wang QJ, Yoon HS (2011) A tunable 3D optofluidic waveguide dye laser via two centrifugal Dean flow streams Lab on a Chip 11:3182-3187 doi:10.1039/c1lc20435a

Young DL, Chou CK, Chen CW, Lai JY, Watson DW (2016) Method of Fundamental Solutions for Three-Dimensional Exterior Potential Flows Journal of Engineering Mechanics 142 doi:10.1061/(asce)em.1943-7889.0001139

Yukselen MA (1996) Superposition technique for two dimensional potential flow around multi-element aerofoils Mechanics Research Communications 23:103-110 doi:10.1016/0093-6413(95)00083-6

Zhang YN, Zhao Y, Zhou TM, Wu QL (2018) Applications and developments of on-chip biochemical sensors based on optofluidic photonic crystal cavities Lab on a Chip 18:57-74 doi:10.1039/c7lc00641a

Zhu XQ, Liang L, Zuo YF, Zhang XM, Yang Y (2017) Tunable Visible Cloaking Using Liquid Diffusion Laser Photon Rev 11 doi:10.1002/lpor.201700066